\documentclass[floatfix,twocolumn,aps,pra,showpacs]{revtex4-1}
\usepackage{graphicx}
\usepackage{amsmath}
\usepackage{amssymb}
\usepackage{soul}

\newcommand{\grad}{\boldsymbol{\nabla}}
\newcommand{\threevec}[3]{\left(\begin{array}{c}#1\\#2\\#3\end{array}\right)}
\newcommand{\bvec}[1]{\mathbf{#1}}
\newcommand{\nematic}{\bvec{\hat{d}}}
\newcommand{\absF}{|\langle\bvec{\hat{F}}\rangle|}
\newcommand{\inleva}[1]{\langle#1\rangle}
\newcommand{\abs}[1]{\left|#1\right|}
\newcommand{\eva}[1]{\left<#1\right>}

\newcommand{\SO}{{\rm SO}}
\newcommand{\U}{{\rm U}}

\newcommand{\beq}{\begin{equation}}
\newcommand{\eeq}{\end{equation}}

\newcommand{\<}{\langle}
\renewcommand{\>}{\rangle}

\renewcommand{\)}{\right)}
\renewcommand{\[}{\left[}
\renewcommand{\]}{\right]}
\newcommand{\rv}{{\bf r}}
\newcommand{\pder}[2]{\frac{\partial#1}{\partial#2}}

\begin{document}

\author{Magnus O.\ Borgh}
\email{M.O.Borgh@soton.ac.uk}
\author{Janne Ruostekoski}
\email{janne@soton.ac.uk}
\affiliation{School of Mathematics, University of Southampton, SO17 1BJ,
    Southampton, UK}

\title{Topological interface physics of defects and textures in spinor
Bose-Einstein condensates}

\begin{abstract}
We provide a detailed description of our previously proposed scheme
for topological interface
engineering with constructed defects and textures perforating across
coherent interfaces between different broken symmetries [M.~O.~Borgh
  and J.~Ruostekoski, Phys.\ Rev.\ Lett.\ \textbf{109,} 015302
  (2012)]. We consider a spin-1 Bose-Einstein condensate,
in which polar and ferromagnetic phases are prepared in spatially
separated regions. We show
that a stable coherent interface is established between the two
phases, allowing
defects of different topology to connect continuously across the
boundary. We provide analytic
constructions of interface-crossing defect solutions that could be
experimentally phase-imprinted using existing technology. By
numerically minimizing the
energy, we calculate the core structures of interface-crossing defect
configurations. We
demonstrate nontrivial core deformations to considerably more complex
structures,  
such as the formation of an arch-shaped half-quantum line defect, an
\emph{Alice arch,} at the 
interface, with the topological charge of a point defect, whose
emergence may be understood by the ``hairy ball" theorem.
Another example of an energetically stable object is the
connection of a coreless vortex to a pair of half-quantum vortices.
We show that rotation leads to spontaneous nucleation of defects in
which a coreless vortex continuously transforms to a half-quantum
vortex across the interface.
\end{abstract}

\maketitle

\section{Introduction}
\label{sec:intro}

In systems described by an order parameter, for example superfluid
liquid helium, liquid crystals and Bose-Einstein condensates (BECs),
the symmetry properties of this order parameter determine the
topological properties of defects and textures~\cite{mermin_rmp_1979}.
Systems with vector order
parameters, such as superfluid liquid $^3$He~\cite{vollhardt-wolfle}
and spinor
BECs~\cite{stenger_nature_1998,kawaguchi_physrep_2012,stamper-kurn_arxiv_2012},
may
exist in several distinct phases. Each phase corresponds to a
different ground-state manifold
of energetically degenerate and physically distinguishable states,
resulting in different order-parameter symmetries.
When two phases coexist in spatially separated regions in the same
medium, topological
defects cannot penetrate the boundary unchanged, but must either
terminate at the interface or connect nontrivially to an object on the other
side. Topological interfaces appear as important objects in seemingly
distant areas of
physics.  For example, it has been
proposed that a series of symmetry breakings in the early universe
leads to the formation of
cosmic strings that terminate on boundaries between regions of
different vacua~\cite{kibble_jpa_1976,vilenkin-shellard}.  Complex
interface physics also arises in string theory from collisions between
branes during inflation~\cite{dvali_plb_1999,sarangi_plb_2002}, in
condensed-matter theory in exotic
superconductivity~\cite{bert_nphys_2011}, and in superfluid
liquid $^3$He when a magnetic-field gradient causes $A$ and $B$ phases to
coexist, resulting in the possibility of nontrivial defects at the phase
boundary~\cite{salomaa_nature_1987,volovik,finne_rpp_2006,bradley_nphys_2008}.
Parallels between defects in superfluids and  objects in
cosmology~\cite{vilenkin-shellard}
led to the suggestion of using
superfluid systems to study analogues of cosmological phenomena in the
laboratory~\cite{volovik,zurek_nature_1985}, including defect formation in
phase transitions
\cite{bauerle_nature_1996,ruutu_nature_1996,weiler_nature_2008},
analogues of brane
annihilation~\cite{bradley_nphys_2008,nitta_pra_2012}, and structures
similar to cosmic
vortons~\cite{battye_prl_2002,savage_prl_2003,ruostekoski_pra_2004}.

Here we present a detailed description of our
proposal for an experimentally feasible scheme to study
topological interface physics in a gas of ultracold atoms with spin degree of
freedom~\cite{borgh_prl_2012}. A stable, coherent interface between
two ground-state
manifolds of different broken symmetries may be prepared by local
manipulation of
the interaction strengths of different scattering channels of the atoms.
The proposed technique can be used to realize, e.g., two different
ground-state manifolds simultaneously in spatially separate regions.
Defects and textures crossing the interface, and connecting objects of
different topologies,
can be created by controlled phase imprinting of elementary
vortex and soliton structures or by rotating the gas. Under
dissipation the phase-imprinted
defect configurations relax to more complex objects for which the
continuous spinor wave function interpolates smoothly across the
topological interface.

As an example we consider defects and textures crossing the boundary
between polar and
ferromagnetic (FM) regions in a spin-1 BEC and show that a coherent
interface is established within a continuous condensate wave function.
We analytically construct
prototype spinor wave functions
representing interface-crossing defect structures, and by numerically
minimizing their energy, we evaluate the configurations that emerge as
a result of energy dissipation in spin-1 BECs.
The simulations demonstrate nontrivial core deformations and defect
structures. We
characterize the defect cores, analyze the energetic stability of
defect solutions crossing the interface, and explain how the
defect-carrying condensate wave function
continuously interpolates across the interface.  We also demonstrate
nucleation of interface-crossing defects consisting of a coreless
vortex that connects to a half-quantum vortex.

A spinor BEC~\cite{stenger_nature_1998,kawaguchi_physrep_2012,stamper-kurn_arxiv_2012}
is created in an all-optical trap so that the spin degree
of freedom of the atoms is not frozen
out by magnetic fields. Spin rotations then
combine with the condensate phase to form a large set of physically
distinguishable states.  Because also the contact interaction between
the atoms in the condensate becomes spin-dependent, energetically
degenerate subsets depend on the strength
and sign of the spin-dependent contributions.  Hence the spinor BEC
exhibits a rich diagram of phases with different broken order-parameter symmetries~\cite{ho_prl_1998,ohmi_jpsj_1998,zhou_ijmpb_2003,koashi_prl_2000,ciobanu_pra_2000,barnett_prl_2006,santos_spin-3_2006}
as a function of the interaction strengths.
This is similar to superfluid liquid $^3$He where
nonzero spin and orbital angular momenta of the Cooper pairs combine
to form phases with different order-parameter
symmetries~\cite{vollhardt-wolfle} supporting a variety of
defects and textures~\cite{salomaa_rmp_1987}.

Modern techniques used in experiments with ultracold atoms
provide tools for unprecedented control over system parameters
and for accurate measurements, including the possibility for \emph{in situ} observation of vortices in spinor BECs. As there has been considerable interest in the studies of the stability properties
of field-theoretical solitons in various physical systems~\cite{manton-sutcliffe,bogomolny_sjnp_1976,jackiw_prd_1976,faddeev_nature_1997},
it is therefore not surprising that this is also followed by an
accelerating theoretical interest in a variety of stable and
meta-stable objects in multicomponent atomic BECs. Perhaps the
simplest of such structures where the multicomponent nature of BECs
plays an important role are 1D vector
solitons~\cite{busch_prl_2001,ohberg_prl_2001, kevrekidis_epjd_2004,
  berloff_prl_2005,shrestha_prl_2009, yin_pra_2011,
  nistazakis_pra_2008,carretero-gonzales_nlin_2008}, typically
consisting of stable combinations of dark and bright solitons in
different condensate components. Higher-dimensional defects and
textures include vortex sheets~\cite{kasamatsu_prl_2003} and 3D
particle-like
solitons~\cite{ruostekoski_prl_2001,alkhawaja_nat_2001,battye_prl_2002,savage_prl_2003,ruostekoski_pra_2004,kawakami_prl_2012}
in two-component (pseudo-spin-1/2) condensates, as well as a rich
phenomenology of defects and textures in
spin-1~\cite{ho_prl_1998,yip_prl_1999,leonhardt_jetplett_2000,stoof_monopoles_2001,isoshima_pra_2002,mizushima_pra_2002,mizushima_prl_2002,martikainen_pra_2002,ruostekoski_monopole_2003,savage_dirac_2003,zhou_ijmpb_2003,reijnders_pra_2004,mueller_pra_2004,saito_prl_2006,ji_prl_2008,takahashi_pra_2009,simula_jpsj_2011,borgh_prl_2012,lovegrove_pra_2012},
spin-2~\cite{semenoff_prl_2007,huhtamaki_pra_2009,kobayashi_prl_2009},
and
spin-3~\cite{santos_spin-3_2006,barnett_pra_2007} BECs.
Interface physics has been studied
in two-component BEC systems, for example in the context of vortex
bifurcation at energetically established interfaces in the phase-separation
regime~\cite{takeuchi_jpsj_2006,kasamatsu_jhep_2010} and interface
collisions~\cite{nitta_pra_2012}.
There is a rapid parallel experimental development, exemplified by
preparation of coreless
vortices and related
textures~\cite{leanhardt_prl_2003,leslie_prl_2009,choi_prl_2012,choi_njp_2012},
as 
well as observations of singular vortices produced in phase
transitions~\cite{sadler_nature_2006}, and of spin-texture
formation~\cite{vengalattore_prl_2008,kronjager_prl_2010,bookjans_prl_2011}. 
Furthermore, trapping of ultracold atoms in artificial gauge-field
potentials~\cite{dalibard_rmp_2011}
was recently realized experimentally~\cite{lin_nature_2009,lin_nature_2011}.
This presents intriguing possibilities for the stability studies of defects and
textures, including those of particle-like solitons~\cite{kawakami_prl_2012}.

Our study of topological interface engineering is organized as follows: In
Sec.~\ref{sec:spin-1-bec}, we first give a brief overview of
the standard mean-field theory of the spin-1 BEC and then proceed to
give a more detailed presentation of the topology and basic defects of the
FM and polar phases. In Sec.~\ref{sec:interface} we discuss how an
interface between FM and polar regions can be created, and then
identify and explicitly construct interface-crossing defect
solutions. We proceed to minimize the energy of the defect solutions
in Sec.~\ref{sec:core-structures} and describe the emerging structures of the
defect core and the energetic stability of the defects.
We explicitly demonstrate continuity of the spinor
wave function across the stable interface.
We summarize our findings in
Sec.~\ref{sec:conclusions}.

\section{Spin-1 BEC}
\label{sec:spin-1-bec}

\subsection{Mean-field theory of spin-1 BEC}
\label{sec:mft}

We consider a trapped
spin-1 atomic BEC confined in an all-optical, harmonic trap.  We may then
employ the classic Gross-Pitaevskii mean-field theory describing a
spatially inhomogeneous macroscopic condensate wave function $\Psi({\bf r})$.
Since we are considering spin-1 atoms, $\Psi({\bf r})$ can be written
in terms of the density of atoms $n(\bvec{r})$ and a
normalized, three-component spinor $\zeta({\bf r})$ in the basis of
spin projection onto the $z$ axis as
\begin{equation}
  \label{eq:spinor}
  \Psi({\bf r}) = \sqrt{n({\bf r})}\zeta({\bf r})
  = \sqrt{n({\bf r})}\threevec{\zeta_+({\bf r})}
                              {\zeta_0({\bf r})}
			      {\zeta_{-}({\bf r})},
  \quad
  \zeta^\dagger\zeta=1.
\end{equation}
The mean-field Hamiltonian density then
reads~\cite{pethick-smith,ho_prl_1998,ohmi_jpsj_1998}
\begin{equation}
  \label{eq:hamiltonian-density}
  \begin{split}
    {\cal H} =  &\frac{\hbar^2}{2m}\abs{\grad\Psi}^2 + V(\bvec{r})n
    + \frac{c_0}{2}n^2
    + \frac{c_2}{2}n^2\abs{\bvec{\eva{\hat{F}}}}^2 \\
    &+ g_1n\eva{\bvec{B}\cdot\bvec{\hat{F}}}
    + g_2n\eva{\left(\bvec{B}\cdot\bvec{\hat{F}}\right)^2}\,.
  \end{split}
\end{equation}
where $V(\bvec{r})$ is an external trapping potential and $m$ is the atomic
mass. In this work we consider the atoms trapped in a slightly
elongated potential, so that
\beq
\label{eq:V}
  V(\rv) = \frac{1}{2}m\omega^2\(x^2+y^2+\frac{z^2}{4}\).
\eeq
The spin operator $\bvec{\hat{F}}$ is given by
a vector of spin-1 Pauli matrices. Its expectation value
$\inleva{\bvec{\hat{F}}}=\zeta_\alpha^\dagger\bvec{\hat{F}}_{\alpha\beta}\zeta_\beta$
is the local spin vector.
A weak external magnetic
field may be imposed, in which case linear and quadratic Zeeman shifts
as described by the last two terms will arise.
Most of our numerical results correspond to cases for which the Zeeman
splitting 
is assumed to be negligible. We note, however, that all our results
remain qualitatively 
the same in the presence of weak Zeeman splitting energy.

We also investigate the configurations of defects and textures in a
rotating trap. In that 
case we minimize the free energy in a rotating frame, corresponding to
\begin{equation}
  \label{eq:free-energy}
  \begin{split}
    H' &= H - \< \mathbf{\Omega}\cdot \mathbf{\hat{L}} \> \\
       &= \int\! d^3 r \[ {\cal H}(\rv) + i\hbar\Omega
       \Psi^\dagger(\rv) \(x\pder{}{y}-y\pder{}{x}\) \Psi(\rv) \],
  \end{split}
\end{equation}
where we have assumed the axis of rotation defined by $\bvec{\Omega}$
to be along the positive $z$ axis, and $\bvec{\hat{{L}}}$ denotes the
angular-momentum operator.

The two interaction terms in Eq.~(\ref{eq:hamiltonian-density}) arise
from the fact that the spins of two colliding spin-1 atoms may
combine to either $0$ or $2$.  There are therefore two $s$-wave
scattering channels, with scattering lengths $a_0$ and $a_2$,
contributing to the contact interaction between
the atoms in the condensate.  Standard angular-momentum
algebra~\cite{pethick-smith} separates the interaction energy into one
spin-independent contribution and one term that depends on the
magnitude of the spin.  The strengths of the spin-independent and
spin-dependent interaction terms are then given by
\beq
\label{eq:c0-c2}
c_0={4\pi\hbar^2(2a_2+a_0)\over 3m}, \quad c_2={4\pi\hbar^2(a_2-a_0)\over 3m},
\eeq
respectively.  Additional magnetic dipole-dipole interactions that may
influence the spin
textures~\cite{vengalattore_prl_2008,simula_jpsj_2011,lovegrove} are
neglected here.

The sign of $c_2$, the strength of the spin-dependent interaction,
determines the magnitude of the spin vector in a uniform ground state,
leading to the two topologically distinct phases of the spin-1 BEC.
If $c_2<0$, energy minimization favors maximized spin magnitude $\absF=1$
in the FM phase.  This is the case
for $^{87}$Rb where $c_0/c_2\simeq -216$~\cite{van-kempen_prl_2002}.
Conversely, if $c_2>0$, as for $^{23}$Na with $c_0/c_2\simeq
31$~\cite{crubellier_epjd_1999}, $\absF=0$ is favored in
the polar phase.  The two phases are described by fundamentally
different order parameters, supporting different families of defects,
which we will discuss in some detail below.

Characteristic length scales arise from the interaction terms.  The
spin-independent interaction defines the usual density healing length
\beq
\label{eq:xi_n}
   \xi_n=\frac{1}{\sqrt{8\pi c_0 n}},
\eeq
which describes the length scale over
which the atom density $n(\bvec{r})$ heals around a local density
depletion.  In
addition, the spin-dependent interaction gives rise to a spin healing
length
\beq
\label{eq:xi_F}
   \xi_F=\frac{1}{\sqrt{8\pi |c_2| n}},
\eeq
defining the distance over which $|\inleva{\bvec{\hat{F}}(\bvec{r})}|$
heals as the order parameter is
excited out of its ground-state manifold.
That situation arises in two cases of importance for the
analysis presented in this article. Firstly, the core of a singular
vortex in one phase may fill with atoms such that the atoms at the
singularity exhibit the opposite phase. This can happen since the
singularity of the spinor order parameter
may be accommodated either by forcing the density to zero or by
requiring that the wave function become orthogonal to the ground-state
manifold (meaning locally perturbing
$\absF$)~\cite{ruostekoski_monopole_2003,lovegrove_pra_2012}. The size
of the filled vortex core is then determined by $\xi_F$.  Secondly, we
are interested here in the interface between polar and FM regions.
For the condensate wave function to interpolate between the two
manifolds, the spin magnitude must leave its ground-state value close
to the interface, which will therefore acquire a width determined by $\xi_F$.

\subsection{Ground state manifolds and basic defects}
\label{sec:topology-defects}

The order parameter manifold is the set of energetically degenerate,
physically distinguishable states.  In the condensation transition,
this symmetry is spontaneously broken, and this broken symmetry
determines the topologically distinct families of defects.  The FM and
polar phases of the spin-1 BEC are described by very
different order-parameter manifolds, leading to dramatically different
possible vortex states.  Before discussing the interface between FM
and polar regions in the next section, we here give an overview of the
families of defects in the purely FM or purely polar BEC.

\subsubsection{The FM phase}

If $c_2<0$ in Eq.~(\ref{eq:hamiltonian-density}), the spin-dependent
interaction
will favor a state that maximizes the magnitude of the spin
everywhere, such that $\absF=1$.  A representative FM spinor is given
by $\zeta = (1,0,0)^T$, such that the spin vector is parallel with the
$z$ axis.  From this representative spinor, a general FM spinor may
be constructed by a 3D spin rotation
\beq
U(\alpha,\beta,\gamma)=\exp(-iF_z\alpha)\exp(-iF_y\beta)\exp(-iF_z\gamma)\,,
\eeq
defined by three Euler angles, together with a condensate phase
$\phi$, as
\begin{equation}
  \label{eq:ferro}
  \zeta^{\rm f} =
  e^{i\phi}U(\alpha,\beta,\gamma)\threevec{1}{0}{0}
  = \frac{e^{-i\gamma^\prime}}{\sqrt{2}}
    \threevec{\sqrt{2}e^{-i\alpha}\cos^2\frac{\beta}{2}}
             {\sin\beta}
             {\sqrt{2}e^{i\alpha}\sin^2\frac{\beta}{2}},
\end{equation}
where the condensate phase is absorbed in the third Euler angle:
\mbox{$\gamma^\prime=\gamma-\phi$}.
Any FM spinor is thus described by some particular choice for
$(\alpha,\beta,\gamma^\prime)$. Therefore the
broken symmetry of the ground-state manifold is represented by the
group of 3D rotations $\SO(3)$. The spin vector is given
by the Euler angles as
\mbox{$\inleva{\bvec{\hat{F}}}=(\cos\alpha\sin\beta,\sin\alpha\sin\beta,\cos\beta)$.}

Topological stability of line defects can be characterized by studying
closed contours around the defect line and the mapping of these contours into
order-parameter space~\cite{mermin_rmp_1979}. If the image in
order-parameter space of a closed
loop encircling a line defect can be contracted to a point, the defect
is not topologically
stable.  The order-parameter space of a FM spin-1 BEC, $\SO(3)$, can
be represented geometrically as $S^3$ (the unit
sphere in four dimensions) with diametrically opposite points
identified.  If a closed contour connects such
identified points more than once, any pair
of such connections can be eliminated by continuous deformation of the
contour.  Therefore any contour with an even number of connections can
be contracted to a point, whereas a contour with an odd number of
connections can be deformed into a contour with just one connection.
Hence we have only two distinct classes of vortices: singly
quantized, singular vortices that correspond to noncontractible loops,
and nonsingular, coreless vortices representing contractible loops. All other
vortices can be transformed to either one of these by local
deformations of the order parameter.  Mathematically, these
equivalence classes are characterized by the two elements of the first
homotopy group, $\pi_1[\SO(3)]=\mathbb{Z}_2$.

The simplest representative of the class of singular line defects is
constructed as a $2\pi$ winding of the condensate phase such that
$\gamma^\prime=-\varphi$ in Eq.~(\ref{eq:ferro}), where
  $\varphi$ is the azimuthal angle.
This results in a line singularity oriented along the $z$ axis.
The spin texture is uniform, such that $\alpha$ and $\beta$ are arbitrary but
constant. The vortex is then described by the spinor
\begin{equation}
  \label{eq:fmsingular}
  \zeta^{\rm s}(\bvec{r}) =
  \frac{e^{i\varphi}}{\sqrt{2}}
  \threevec{\sqrt{2}e^{-i\alpha}\cos^2\frac{\beta}{2}}
           {\sin\beta}
           {\sqrt{2}e^{i\alpha}\sin^2\frac{\beta}{2}}.
\end{equation}

\begin{figure}[tb]
  \centering
  \includegraphics[scale=1]{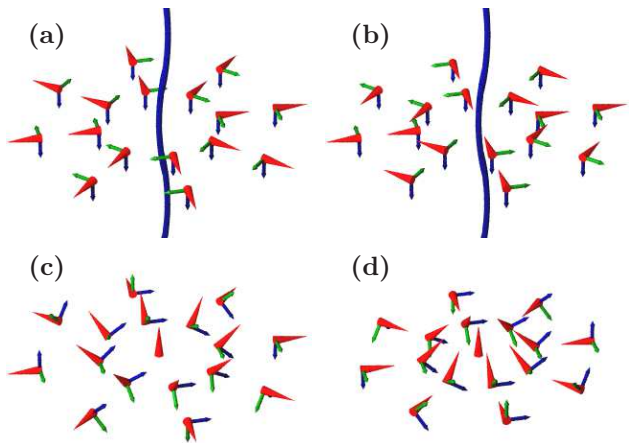}
\caption{(Color online) Nontrivial vortices in the FM phase.
  (a) A radial disgyration of
  the spin vector (red cones) around the singular core represents a
  spin vortex.  This vortex carries a spin current but no mass
  circulation, a manifestation of the nonquantization of circulation
  alone in the FM phase.
  (b) All vortices in the class of singly quantized, singular vortices
  can be deformed into each other by local spin rotations. For
  example, it is possible to form a  spin vortex with a
  cross-disgyration of the spin vector.
  (c) In addition to the one class of singular vortices, the FM phase
  also supports nontrivial, nonsingular vortices.  In a coreless
  vortex the order parameter remains nonsingular everywhere in the
  vortex structure.  The characteristic fountain structure of the spin
  is formed by a rotation of the spin vector around the vortex line,
  together with a winding of
  the condensate phase, corresponding to spin rotations about the
  local spin vector (indicated by the orthogonal green and blue
  vectors).
  (d) Like the class of singular line defects, members of the  family of
  nonsingular vortices are related by local deformations of the order
  parameter allowing different vortex configurations.
}
\label{fig:fmVortices}
\end{figure}

From $\zeta^{\rm s}$ other vortices in the same equivalence class can
be formed by local spin rotations. For example, we may rotate the
spin vector such that at each point it points radially away from the
vortex line, as illustrated in Fig.~\ref{fig:fmVortices}(a).  This
vortex corresponds to the choices $\alpha=\varphi$, $\beta=\pi/2$ and
$\gamma^\prime=0$ in Eq.~(\ref{eq:ferro}), and is described by
\begin{equation}
  \label{eq:spinvortex}
  \zeta^{\rm sv} =
  \frac{1}{2}\threevec{e^{-i\varphi}}
                      {\sqrt{2}}
                      {e^{i\varphi}}.
\end{equation}
This singular spin vortex~\cite{ho_prl_1998,ohmi_jpsj_1998}
illustrates another important aspect of the FM phase: circulation
alone is not quantized.  The superfluid velocity in the FM
phase~\cite{ho_prl_1998},
\begin{equation}
  \label{eq:fm-velocity}
  \bvec{v} =
  -\frac{\hbar}{m}(\grad\gamma^\prime+\cos\beta\grad\alpha),
\end{equation}
vanishes when $\gamma^\prime=0$ and $\beta=\pi/2$.
Therefore the circulation is in fact zero in
$\zeta^{\rm sv}$, but it does carry a nonvanishing spin current around a
singularity of the FM spin vector, whose structure is similar to an
analogous vortex with a radial disgyration of the angular momentum
vector in $^3$He.
Further local spin rotations yield other singular vortices with
different spin structures, such as the cross disgyration shown in
Fig.~\ref{fig:fmVortices}(b) or a tangential disgyration with
$\inleva{\bvec{\hat{F}}}=\boldsymbol{\hat{\varphi}}$.

A striking manifestation of the nonquantization of circulation in the
FM phase is the formation of a nonsingular coreless vortex.  This can
be constructed as a combined rotation of the spin vector and
the condensate phase [Fig.~\ref{fig:fmVortices}(c)]:
$\alpha=-\gamma^\prime=\varphi$, yielding the spinor
\begin{equation}
  \label{eq:cl}
  \zeta^{\rm cl}(\bvec{r}) =
  \frac{1}{\sqrt{2}}\threevec{\sqrt{2}\cos^2\frac{\beta(\rho)}{2}}
                          {e^{i\varphi}\sin\beta(\rho)}
                          {\sqrt{2}e^{2i\varphi}\sin^2\frac{\beta(\rho)}{2}},
\end{equation}
where the Euler angle $\beta$ is now a function of the radial distance
$\rho=\sqrt{x^2+y^2}$, such that $\beta \to 0$ as $\rho \to 0$,
keeping the spin texture continuous.  The superfluid
velocity, Eq.~(\ref{eq:fm-velocity}), becomes
\begin{equation}
  \label{eq:cl-velocity}
  \bvec{v}^{\rm cl} =
  \frac{\hbar}{m\rho}(1-\cos\beta)\boldsymbol{\hat{\varphi}},
\end{equation}
and increases smoothly from zero at $\rho=0$ as $\beta$ increases away
from the vortex, the spin vector forming a fountain-like
texture. The coreless vortex in the FM phase of a spin-1 BEC is
analogous to the Anderson-Toulouse and Mermin-Ho vortices in superfluid
$^3$He~\cite{anderson_prl_1977,mermin_prl_1976}, which differ by
the boundary condition imposed on the angular momentum vector at the
container wall.  In the BEC there is no hard container wall, and the
amount by which $\beta$ turns from the vortex line to the edge of the
cloud is determined by the rotation of the trap, causing the total
angular momentum to vary smoothly with rotation.

The coreless vortex can be continuously transformed into other members
of the class of nonsingular vortices, including the vortex-free
state, by purely local operations. The continuous deformation is a striking consequence of the
two-element character of the
fundamental group of the $\SO(3)$ order-parameter space: the doubly
quantized vortex belongs to the same topological class as the
nonsingular vortices and the vortex-free state, and can be
continuously unwound, if the orientation of the spin texture is not
fixed outside the structure. Another
nontrivial nonsingular vortex with continuous spin textures is displayed
in Fig.~\ref{fig:fmVortices}(d).

We have now identified two topologically distinct classes of vortices
that can both carry mass and spin circulation in the FM spin-1 BEC.  A
similar situation applies in the $A$ phase of $^3$He. There the singular
vortex has the lower energy, but the energy barrier for nucleation is
lower for the nonsingular vortex~\cite{parts_prl_1995}.
In a rotating FM spin-1 BEC, the coreless vortex has the lower energy
and the lower nucleation barrier, and consequently the ground
state is made up of nonsingular coreless vortices for sufficiently
rapid rotation~\cite{mizushima_prl_2002, martikainen_pra_2002,
  reijnders_pra_2004, mueller_pra_2004, takahashi_pra_2009}.  However,
it is also possible to form a singly quantized, singular
vortex~\cite{isoshima_pra_2002,mizushima_pra_2002}, which
despite not being the lowest-energy state at any frequency of
rotation can nevertheless be energetically stable as a local energy
minimum~\cite{lovegrove_pra_2012}.

\begin{figure}[tb]
  \centering
  \includegraphics[scale=1]{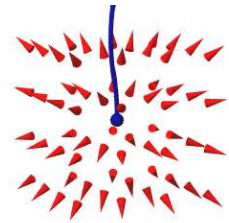}
\caption{(Color online) An analogue of the field-theoretical Dirac
  monopole can be
  constructed in the FM phase of the spin-1 BEC.  The radial hedgehog
  monopole structure of the spin vector (cones) appears as the termination of
  a doubly quantized, singular vortex line.}
\label{fig:dirac}
\end{figure}

So far we have considered line defects, classified by the first
homotopy group $\pi_1$.  Point defects---monopoles---are analogously
classified by the second homotopy group $\pi_2$.  For the FM order-parameter manifold, the
second homotopy group is the trivial group, $\pi_2[\SO(3)]=0$, indicating
that the FM phase does not strictly speaking support point defects.
However, it is possible to form a spinor with a
monopole structure of the spin vector (a radial hedgehog) as the
termination of a doubly quantized vortex~\cite{savage_dirac_2003}
(Fig.~\ref{fig:dirac}).
This is the analogue of the Dirac monopole in quantum field theory,
and the doubly quantized vortex line is called the Dirac string.  The
corresponding spinor is written by choosing
$\alpha=\gamma^\prime=\varphi$ and $\beta=\theta$ (where $\theta$ and
$\varphi$ are the polar and azimuthal angles, respectively) to form
\begin{equation}
  \label{eq:dirac}
  \zeta^{\rm D} = \frac{1}{\sqrt{2}}
                  \threevec{\sqrt{2}e^{-2i\varphi}\cos^2\frac{\theta}{2}}
                   {e^{-i\varphi}\sin\theta}
                   {\sqrt{2}\sin^2\frac{\theta}{2}}.
\end{equation}
This Dirac monopole can be continuously deformed into the spin
structure of the coreless vortex~\cite{savage_dirac_2003}.

\subsubsection{Polar phase}

We next consider $c_2>0$ in
  Eq.~(\ref{eq:hamiltonian-density}), in which case the
spin-dependent interaction
favors a state with $\absF=0$.  A simple
representative polar spinor fulfilling this requirement is
$\zeta=(0,1,0)^T$.  As for the FM phase, the general polar spinor is
found by applying a 3D spin rotation $U(\alpha,\beta,\gamma)$ together
with a condensate phase $\phi$:
\begin{equation}
  \label{eq:polar}
  \zeta^{\rm p} =
  e^{i\phi}U(\alpha,\beta,\gamma)\threevec{0}{1}{0}
  = \frac{e^{i\phi}}{\sqrt{2}}\threevec{-e^{-i\alpha}\sin\beta}
                     {\sqrt{2}\cos\beta}
                     {e^{i\alpha}\sin\beta}.
\end{equation}
We now make the important observation that the unit vector
$\nematic = (\cos\alpha\sin\beta,\sin\alpha\sin\beta,\cos\beta)$
defines the local direction of macroscopic condensate spin
quantization.  This allows us to rewrite $\zeta^{\rm p}$ in terms of
this vector as~\cite{ruostekoski_monopole_2003}
\begin{equation}
\label{eq:nematic}
  \zeta^{\rm p} = \frac{e^{i\phi}}{\sqrt{2}}
                 \threevec{-d_x+id_y}{\sqrt{2}d_z}{d_x+id_y}.
\end{equation}
The condensate phase $\phi$, which takes values on a unit circle, and
the unit vector $\nematic$, taking values on a sphere, thus together
fully specify the order parameter in the polar phase. Note, however, that
$\zeta^{\rm p}(\phi,\nematic) = \zeta^{\rm p}(\phi+\pi,-\nematic)$.
These two states must be identified in order to avoid double counting.
The order parameter space therefore becomes
$(\U(1) \times S^2)/\mathbb{Z}_2$, where the factorization by the
two-element group $\mathbb{Z}_2$ results from the identification.
The vector $\nematic$ should thus be taken to be \emph{unoriented} and defines
a \emph{nematic axis}~\cite{zhou_ijmpb_2003}, and the order-parameter
is correspondingly said to exhibit \emph{nematic order,} which leads
to parallels with the $A$-phase of superfluid $^3$He.

A simple singly quantized vortex can again be constructed as a $2\pi$ winding
of the condensate phase, keeping $\nematic$ uniform (choosing $\alpha$
and $\beta$ to be constants):
\begin{equation}
  \label{eq:psingular}
  \zeta^{\rm 1} =
  \frac{e^{i\varphi}}{\sqrt{2}}\threevec{-e^{-i\alpha}\sin\beta}
                     {\sqrt{2}\cos\beta}
                     {e^{i\alpha}\sin\beta}.
\end{equation}
In the polar phase the superfluid velocity is~\cite{kawaguchi_physrep_2012}
\begin{equation}
  \label{eq:polar-velocity}
  \bvec{v} = \frac{\hbar}{m}\grad\phi.
\end{equation}
We observe that $\bvec{v}$ depends
only on the gradient of the condensate phase, and is independent of
$\nematic$.  This means that another singly quantized vortex, with the same
circulation as that described by Eq.~(\ref{eq:psingular}), can be
formed by allowing $\nematic$ to wind by $2\pi$ (thus preserving
single-valuedness of the order parameter) in addition to the winding
of the condensate phase.  This is achieved by choosing
$\alpha=\phi=\varphi$ in Eq.~(\ref{eq:polar}), yielding the spinor
\begin{equation}
  \label{eq:p012}
  \zeta^{\rm 1^\prime} =
  \frac{1}{\sqrt{2}}\threevec{-\sin\beta}
                             {\sqrt{2}e^{i\varphi}\cos\beta}
                             {e^{2i\varphi}\sin\beta}.
\end{equation}

\begin{figure}[tb]
  \centering
  \includegraphics[scale=1]{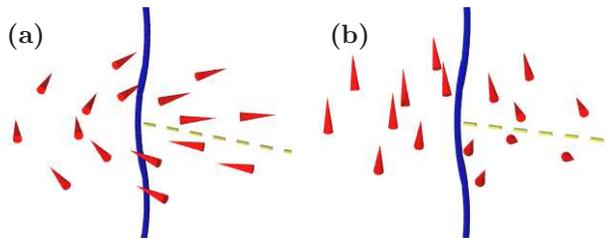}
\caption{(Color online) The polar phase of the spin-1 BEC supports half-quantum
  vortices, constructed as a $\pi$ winding of the condensate phase
  (which determines the quantized circulation) together with a
  $\nematic\to-\nematic$ winding of the nematic axis. The
  identification $\zeta^{\rm p}(\phi,\nematic) = \zeta^{\rm
  p}(\phi+\pi,-\nematic)$ keeps the order parameter single-valued.
  (a) Half-quantum vortex as described by Eq.~(\ref{eq:hq}).
  The nematic axis (red cones) winds by $\pi$ in the plane
  perpendicular to the vortex line as the core is
  encircled.  The disclination plane where the
  $\nematic\leftrightarrow-\nematic$ identification is made is
  indicated by the dashed line.
  (b) The winding of the nematic axis need not stay in the plane
  perpendicular to the vortex line.  For example, a half-quantum
  vortex can also be constructed where $\nematic$ winds in a plane
  parallel to the vortex line as the defect is encircled.
}
\label{fig:hq}
\end{figure}

One can further show from Eq~(\ref{eq:polar-velocity}) that
circulation is quantized in the polar phase.  However, due to the
nematic order, the smallest circulation possible is half that of a
singly quantized vortex.  The equivalence
$\zeta^{\rm p}(\phi,\nematic) = \zeta^{\rm p}(\phi+\pi,-\nematic)$
implies that we can allow the condensate phase to wind by $\pi$ along
a loop encircling the vortex and still preserve single-valuedness of
the the order parameter by a simultaneous $\nematic \to -\nematic$
winding of the nematic axis~\cite{leonhardt_jetplett_2000}.
If $\nematic$
is in the $(x,y)$-plane, a half-quantum vortex can be
written
\begin{equation}
  \label{eq:hq}
  \zeta^{\rm hq} = \frac{e^{i\varphi/2}}{\sqrt{2}}
                   \threevec{-e^{-i\varphi/2}}
                            {0}
                            {e^{i\varphi/2}}
		 = \frac{1}{\sqrt{2}}
                   \threevec{-1}
                            {0}
                            {e^{i\varphi}}.
\end{equation}
In general, the axis about which $\nematic$ winds need not coincide
with the vortex core.  Figure~\ref{fig:hq}(b) shows a half-quantum vortex
where $\nematic$ winds about an axis perpendicular to the vortex
line. This vortex is related to that shown in Fig~\ref{fig:hq}(a) and defined by
Eq.~(\ref{eq:hq}) by a spin rotation.  The resulting spinor wave
function may appear quite complicated, but the $\pi$ winding of the
nematic axis still
allows us to identify the vortex.

Thus circulation is quantized in the polar phase, and
indeed one can show that $\pi_1[(\U(1) \times
  S^2)/\mathbb{Z}_2]=\mathbb{Z}$. The topological charges
$1/2$, $1$, $3/2$ etc.\ are additive.  For example, the state with two
half-quantum vortices belongs to the same equivalence class as the
singly quantized vortices.  This observation shall prove important for
understanding the core structure of defects that cross a polar-FM
interface.

In addition to singular line defects, the polar phase also supports
singular point
defects~\cite{stoof_monopoles_2001,ruostekoski_monopole_2003}:
A spherically symmetric point defect, analogous to the
't~Hooft-Polyakov monopole in quantum field theory, is formed by
choosing $\nematic = \bvec{\hat{r}} =
(\sin\theta\cos\varphi,\sin\theta\sin\varphi,\cos\theta)$ in
Eq.~(\ref{eq:polar}), such that the vector field $\nematic$ forms a
radial hedgehog structure
\begin{equation}
  \label{eq:polar_monopole}
  \zeta^{\rm pm} = \frac{1}{\sqrt{2}}\threevec{-e^{-i\varphi}\sin\theta}
                            {\sqrt{2}\cos\theta}
	                    {e^{i\varphi}\sin\theta}.
\end{equation}
Here a singular point defect is located at the origin. The two spinor
wave function components $\zeta_\pm$
form overlapping, singly quantized vortex lines with opposite
circulation. The vortex lines are oriented along the $z$ axis, normal
to a dark soliton plane (phase kink) in the component $\zeta_0$.

\section{Topological interface in a spin-1 BEC}
\label{sec:interface}

The two phases of the spin-1 BEC exhibit different topological
properties of the order parameter, which is manifest in the very
different defects they support, as detailed in the previous section.  We
now consider the behavior of the
order parameter and defects when the two phases are realized
simultaneously in spatially separated regions within the same ultracold
gas, so that an interface between the different ground-state manifolds
is formed.

The order-parameter manifolds in the FM and polar phases emerge out of
the full symmetry of the condensate wave function as the
spin-dependent interaction selects an energetically degenerate subset
of all possible wave functions.  In particular, the order-parameter
space in the FM phase consists of all spinor wave functions that maximize
the spin magnitude, $\absF=1$, everywhere if the texture is
uniform.  Correspondingly in the polar phase, the order-parameter
space is the set of wave functions that have $\absF=0$ everywhere.
These sets are clearly nonoverlapping.  However,
because they form subsets of the same spin-1 wave
function, a continuous connection between spatially separated polar
and FM manifolds is possible
by exciting the wave function out of its ground-state manifold close
to the interface, locally restoring its full symmetry.

In this section we first discuss how the interface may be created in
the spinor BEC through local manipulation of the scattering lengths.
We then identify the basic interface-crossing defect states and
explicitly construct prototype spinor wave functions to describe them.  In
Sec.~\ref{sec:core-structures} we minimize the energy of these
spinor wave functions for defect configurations and show
how this leads to a rich phenomenology of defect structures.

\subsection{Creating the topological interface}
\label{sec:creating-interface}

In order to realize a topological interface in a spinor BEC it is
necessary make the system switch continuously between regions of
different broken symmetries. Which broken symmetry is preferred in a
spin-1 BEC on 
energetic grounds is determined, as explained in Sec.~\ref{sec:mft},
by the spin-dependent interaction. Therefore if  one can spatially
control the interaction strength $c_2$, Eq.~(\ref{eq:c0-c2}), separate
FM and polar regions within
the same BEC can be engineered.  Specifically, since $c_2 \propto
(a_2-a_0)$, this implies changing the ratio $a_0/a_2$ of the two
scattering lengths of colliding spin-1 atoms such that $c_2$ changes
sign.

The scattering lengths that determine interaction strengths in
ultracold-atom systems are routinely manipulated
using magnetic Feshbach resonances.  However, this technique cannot be
used for our present purpose since the strong magnetic
fields required would freeze out the spin degree of
freedom and destroy the spinor nature of the BEC.
The possibility for engineering the scattering lengths in the
spinor BEC is instead provided by the use of either
optical~\cite{fatemi_prl_2000} or
microwave-induced Feshbach
resonances~\cite{papoular_pra_2010}, in which case the fields can be
kept sufficiently
weak in order not to destroy the spinor nature of the BEC.

We suggest constructing
an interface between topologically distinct manifolds in a spinor BEC
by local adjustment of the scattering lengths, such that regions with
different-sign $c_2$ are created~\cite{borgh_prl_2012}.
In a spin-1 BEC experiment, the spatial dependence of the scattering
lengths can then result in an interface between coexisting FM and
polar phases.  Doing so, however, presents practical challenges. If an optical
\mbox{Feshbach} resonance is used to manipulate one or both scattering
lengths, the spatial pattern corresponding to a sharp
interface can be imposed using a
holographic mask.  Optical Feshbach resonances suffer from
inelastic losses~\cite{fatemi_prl_2000}, but these can be kept small
for small adjustments of the scattering lengths.  Since the
spin-dependent interaction is proportional to the difference between
$a_0$ and $a_2$, only a small relative shift is needed to create the
interface if $|c_2|$ is small, which is true for both $^{87}$Rb and
$^{23}$Na, commonly used in spinor-BEC experiments.

Using a microwave-induced Feshbach resonance avoids the problem of
large inelastic losses, but makes engineering the spatial profile more
difficult, since, except in specific traps, for example surface microtraps,
a microwave field cannot be focused in the same way as
the laser.  This problem could be overcome by using an optically
induced level shift to tune the microwave transition off-resonant.
The microwave field could then be applied uniformly across the system,
whereas spatial control of the laser field is used to apply the
optical tuning only in the region where no shift of $c_2$ is required.

\subsection{Construction of prototype interface spinors}
\label{sec:defect-solutions}

The use of optical or microwave-induced Feshbach resonances
can thus realize spatially separated polar and FM phases in the same
spin-1 BEC,
with the condensate wave function remaining continuous, allowing, in
principle, defects to connect across the interface.
In order to demonstrate the nontrivial nature of defect penetration
across the interface
between topologically distinct manifolds, we consider a spin-1 BEC
where $c_2$ abruptly changes sign at $z=0$. We choose $c_2>0$ for
$z>0$ and $c_2<0$ for $z<0$, such that the interface exists at $z=0$
with the polar phase above it and the FM phase below.  In the
following, we analytically
construct spinor solutions that represent physical wave
functions for defects and textures simultaneously in the two different
manifolds.

\begin{figure}[tb]
  \centering
  \includegraphics[width=0.98\columnwidth]{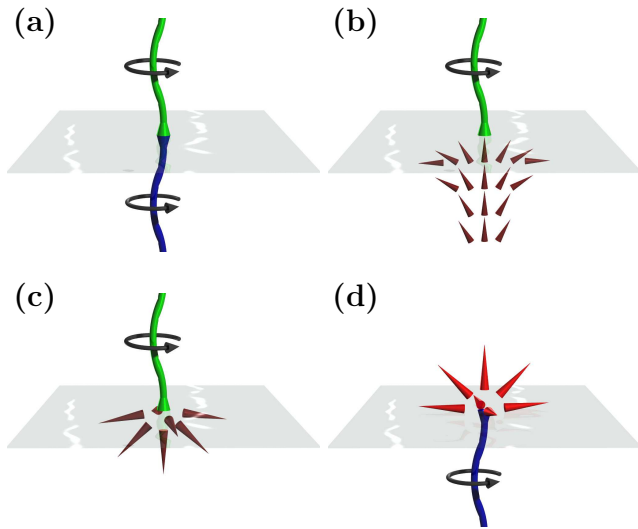}
  \caption{(Color online)
    Schematic illustrations of possible vortex connections. The polar phase
    is above the interface and the FM phase
    below.
    (a) The simplest vortex connection to construct is to consider
    a singly quantized vortex in either phase.  Even though these are
    different objects, representing different topologies, their wave
    functions can be joined continuously across the interface.
    (b) If the singly quantized vortex in the polar phase also
    contains a $2\pi$ winding of the nematic axis, its spinor wave
    function can be made to connect to a spinor representing a coreless
    vortex in the FM phase.
    (c) The FM Dirac monopole has a spinor structure similar to the
    coreless vortex (in fact, the monopole can be continuously unwound
    into a coreless vortex~\cite{savage_dirac_2003}) and it is
    therefore possible for it to
    form the termination point of a singly quantized polar vortex that
    ends at the interface.
    (d) Similarly, a polar monopole on the interface can form the
    termination point of a singly quantized FM spin vortex.  The
    spinor is constructed by noticing that both defects have common
    phase windings of the spinor components.}
  \label{fig:schematic1}
\end{figure}

The simplest vortex connection can be identified by considering a
singly quantized vortex in both phases, as illustrated schematically
in Fig.~\ref{fig:schematic1}(a).  Note that a singly quantized vortex
does not mean the same thing in the two phases: the topology that
describes vortices is entirely different, one vortex being a product
of the broken symmetry 
manifold SO(3), with the fundamental homotopy group of two elements,
and the other one resulting from the broken symmetry
$(\U(1) \times S^2)/\mathbb{Z}_2$,
with the fundamental homotopy group
of integers that represent the number of half-quanta of circulation.
It is therefore 
not obvious that these different topological objects can be
continuously joined across the interface.

In the following we show how to construct spinor wave-function
solutions that simultaneously
represent a singly quantized vortex line in both phases and perforate
through the interface 
with a $2\pi$ winding of the condensate phase around the vortex line
(which we take to be along 
the $z$ axis). A similar procedure is
then extended to other topological defects and textures.
The joining of two singly quantized vortex lines can be achieved by
changing the sign of either 
of the spinor components $\zeta_+$ or $\zeta_-$. By appropriate choice
of parameters doing so 
causes the spinor wave function to adjust between the two manifolds by
forcing $\absF$ to switch from 0 to 1, or else leads to a state which
immediately relaxes to the desired configuration.
Physically, such a sign-change in one
of the two spinor components can be obtained by introducing a dark
soliton plane (phase kink) in that component at $z=0$. The $\pi$ phase
shift across the soliton is then associated with a
vanishing density in that spinor component at the soliton core.  The
BEC wave function, however, remains continuous across the interface, since the
remaining spinor components have nonvanishing atom densities also at
the position of the soliton plane. The BEC wave function thus connects the two
manifolds. In this construction, the switch between polar and
FM sides is abrupt. In section~\ref{sec:core-structures} we will
see that as energy is relaxed, the interface acquires a finite width
determined by the spin healing length $\xi_F$, Eq.~(\ref{eq:xi_F}).

Following this procedure and
starting from the expression for a singular vortex in the FM phase,
Eq.~(\ref{eq:fmsingular}), we
can write the spinor wave function connecting two singly quantized vortices across
the polar-FM interface explicitly as
\begin{equation}
  \label{eq:1qs}
  \zeta^{\rm 1\leftrightarrow s} =
  \frac{e^{i\varphi}}{\sqrt{2}}
  \threevec{\sqrt{2}e^{-i\alpha}\cos^2\frac{\beta}{2}}
           {\sin\beta}
           {\mp\sqrt{2}e^{i\alpha}\sin^2\frac{\beta}{2}},
\end{equation}
where the negative sign is used on the polar side of the
interface and the positive sign on the FM side.  Note that only the
choice $\beta=\pi/2$ yields
$|\inleva{\bvec{\hat{F}}}|=0$ corresponding to an exactly polar state
above the interface.  However, even for a different $\beta$ the spinor
wave function has the appropriate vortex structure and will quickly
relax to the polar phase for $z>0$ with a singly quantized vortex.  This
highlights the general consideration that even though writing exact
vortex connections analytically may be very complicated, we have a
simple method for finding approximate spinor wave functions representing
defect connections.

In Eq.~(\ref{eq:p012}) we demonstrated that a singly quantized vortex
in the polar phase can include a $2\pi$ winding of the nematic axis
in addition to the $2\pi$ winding of the condensate phase.  Comparing
Eq.~(\ref{eq:p012}) with the solution for a coreless FM vortex in
Eq.~(\ref{eq:cl}),
we note that these have a
similar structure in terms of the complex phases of the spinor
components.  We can therefore construct an approximate wave function
representing the connection of a singly quantized vortex in the polar
phase with a coreless vortex on the FM side of the interface by the
insertion of a soliton plane in $\zeta_-$ in Eq.~(\ref{eq:p012}). This
yields the interface spinor
\begin{equation}
  \label{eq:s-cl}
  \zeta^{{\rm 1}\leftrightarrow{\rm cl}} =
  \frac{1}{\sqrt{2}}\threevec{-\sin\beta}
                             {\sqrt{2}e^{i\varphi}\cos\beta}
                             {\pm e^{2i\varphi}\sin\beta},
\end{equation}
where the positive sign is used on the polar side of the
interface and the negative sign on the FM side. Choosing $\beta=\pi/4$ or
$\beta=3\pi/4$ yields $|\inleva{\bvec{\hat{F}}}|=1$ on the FM
side, and specifically the choice $\beta=3\pi/4$
approximates the coreless vortex $\zeta^{\rm cl}$ [Eq.~(\ref{eq:cl})].
This solution relaxes to the characteristic fountain-like spin profile
to yield the
state illustrated in Fig.~\ref{fig:schematic1}(b).

In addition to connecting to another vortex across the interface, a
vortex could also terminate on a point defect at the
interface.  Such solutions can be constructed by joining the
monopole spinor wave functions of Eqs.~(\ref{eq:dirac}) and
(\ref{eq:polar_monopole}) to vortices with analogous
phase windings in each of the spinor components.  For
example, consider the Dirac monopole in the FM phase
[Eq.~(\ref{eq:dirac})].
By the same construction that resulted in the interface-crossing defect
in Eq.~(\ref{eq:s-cl}) we can connect the singular vortex of
Eq.~(\ref{eq:p012})
on the polar side to the monopole of Eq.~(\ref{eq:dirac}) at the
interface by inserting a soliton plane into $\zeta^{\rm D}_-$. The
resulting spinor wave function,
\begin{equation}
  \label{eq:s-dirac}
  \zeta^{\rm 1 \leftrightarrow D} = \frac{1}{\sqrt{2}}
                  \threevec{\sqrt{2}e^{-2i\varphi}\cos^2\frac{\theta}{2}}
                           {e^{-i\varphi}\sin\theta}
                           {\mp\sqrt{2}\sin^2\frac{\theta}{2}}\,,
\end{equation}
represents the monopole on the FM side of the interface. Here the
negative sign refers
to the polar side and the positive sign to the FM side. On the
polar side the spinor has a structure similar to
Eq.~(\ref{eq:p012}), thus approximating a singly quantized vortex in
the polar phase.  The resulting defect configuration is illustrated in
Fig.~\ref{fig:schematic1}(c) and represents a vortex in the
polar phase terminating to a monopole at the interface.
This defect is closely related to the one shown in
Fig.~\ref{fig:schematic1}(b),
as the Dirac monopole and the coreless
vortex can be deformed into each other by purely local operations.
Note also that in Eq.~(\ref{eq:s-dirac}) the Dirac string is represented by
the singular polar vortex along the positive $z$ axis and there is no
line defect on the FM side. By instead aligning the Dirac string with
the negative  $z$ axis, the doubly quantized line defect terminates on
the monopole from the FM side, while for positive $z$, the spinor
still represents a singular vortex connecting to the monopole from the
polar side.

In a similar way a point defect (radial hedgehog) in the polar side
[Eq.~(\ref{eq:polar_monopole})] can be placed
on the interface as the termination point of a singular FM vortex.
We consider a defect structure with overlapping, singly quantized
vortex lines in $\zeta_\pm$, both oriented normal to the interface
and of opposite circulation, together with $\pi$ phase kinks in $\zeta_+$
and $\zeta_0$. This spinor wave function can be parametrized as
\begin{equation}
\label{eq:sv-pm}
  \zeta^{\rm sv\leftrightarrow pm} =
  \frac{1}{\sqrt{2}}\threevec{\mp e^{-i\varphi}\sin\theta}
                             {\sqrt{2}\cos\theta}
	                     {e^{i\varphi}\sin\theta}\,,
\end{equation}
using the negative sign on the polar side and the positive sign on the
FM side.
The resulting structure on the polar side is that of
Eq.~(\ref{eq:polar_monopole}), in which the nematic axis $\nematic$
forms a radial hedgehog
$\nematic=\bvec{\hat{r}}$~\cite{stoof_monopoles_2001,ruostekoski_monopole_2003}.
This represents the polar point defect on the interface. On the
FM side the spinor is similar to the singular spin vortex $\zeta^{\rm sv}$ of
Eq.~(\ref{eq:spinvortex}) with vortex lines
of opposite winding in $\zeta_\pm$. Hence we have constructed on the
FM side an approximation to a spin vortex that terminates to
the polar monopole at the interface, as illustrated in
Fig.~\ref{fig:schematic1}(d).

\begin{figure}[tb]
  \centering
  \includegraphics[width=0.98\columnwidth]{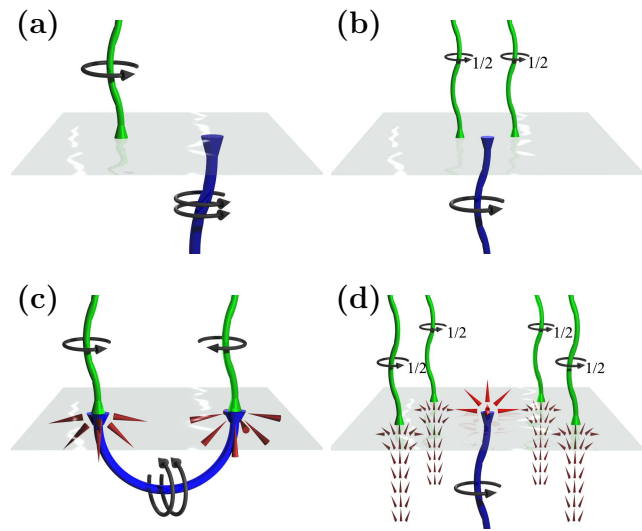}
  \caption{(Color online)
    Schematic illustrations of more complicated vortex connections
    that can be constructed from the basic defect solutions
    illustrated in Fig.~\ref{fig:schematic1}.
    (a) A singly quantized vortex in the polar phase connecting to a
    doubly quantized vortex on the FM side may be cut in half at the
    interface and the resulting vortices
    in the two regions may be moved apart if
    an additional dark soliton plane is introduced in $\zeta_0$.
    (b) A singly quantized polar vortex can split into two
    half-quantum vortices when its energy
    relaxes~\cite{lovegrove_pra_2012}. A repulsive force between the
    half-quantum vortices makes the splitting energetically favorable.
    The splitting mechanism can then yield a state where a FM
    vortex connects to a pair of half-quantum vortices.
    (c) A Dirac dipole can be constructed by joining the Dirac strings
    of a Dirac
    monopole and an antimonopole~\cite{savage_dirac_2003}.  Placed on
    the interface, the dipole
    connects to two singly quantized vortices on the polar side.
    (d) Rotation may cause interface-crossing vortex structures to
    nucleate.  
    Here four nucleated vortex complexes are drawn together
    with a singular vortex already present.}
  \label{fig:schematic2}
\end{figure}

Next we show that vortices can also be made to terminate at the interface.
In Eq.~(\ref{eq:s-cl}) and in Fig.~\ref{fig:schematic1}(b), a singular,
singly quantized polar vortex perforates the interface to a coreless
FM vortex when a $\pi$ phase kink is inserted in $\zeta_-$. The
resulting defect can be cut in half
while still preserving the coherent interface with a continuous
order-parameter field by inserting an additional phase kink in $\zeta_0$.
This allows the vortices on different sides of the interface to move apart:
\begin{subequations}
\begin{align}
    \zeta^{\rm cut} &=
    \frac{1}{\sqrt{2}}\threevec{-\sin\beta}
                               {\sqrt{2}e^{i\varphi}\cos\beta}
                               {e^{2i\varphi}\sin\beta},
    \quad {\rm for}\,\,\, z>0 \\
    \zeta^{\rm cut} & =
    \frac{-1}{\sqrt{2}}\threevec{\sin\beta}
                         {\sqrt{2}e^{i\varphi}\cos\beta}
	  		 {e^{2i\varphi}\sin\beta}, \quad {\rm for}\,\,\, z<0,
\end{align}
\end{subequations}
where we may choose $\beta=3\pi/4$ as in Eq.~(\ref{eq:s-cl}).
One possible configuration is illustrated in Fig.~\ref{fig:schematic2}(a), where
the singular polar vortex and a doubly quantized FM vortex are spatially
separated and both terminate on the interface.

Since the vortex lines in the individual spinor components terminate on the
soliton planes, it is also possible to consider a state where a vortex exists
only on one side of the interface, for instance,
\begin{subequations}
\begin{align}
    \zeta^{\rm pv}  & =
    \frac{1}{\sqrt{2}}\threevec{-\sin\beta}
                               {\sqrt{2}e^{i\varphi}\cos\beta}
                               {e^{2i\varphi}\sin\beta},
    \quad {\rm for}\,\,\, z>0 \\
    \zeta^{\rm pv} & =
    -\frac{1}{\sqrt{2}}\threevec{\sin\beta}
                               {\sqrt{2}\cos\beta}
			       {\sin\beta}, \quad {\rm for}\,\,\, z<0.
\end{align}
\end{subequations}

In addition to singly quantized vortices also half-quantum vortices are
possible in the polar phase [Eq.~(\ref{eq:hq})]. A singly quantized
vortex can split into two half-quantum vortices while
preserving the topology, and such splitting can be energetically
favorable~\cite{lovegrove_pra_2012}. The defect configuration of perforating
singly quantized vortices in Fig.~\ref{fig:schematic1}(a) can therefore also
deform to a state in which a singly quantized FM vortex continuously connects
to a pair of half-quantum vortices as illustrated in
Fig.~\ref{fig:schematic2}(b).  We will demonstrate in the next section
that this state does indeed appear as a consequence of energy
minimization.

The vortices and monopoles in the spin-1 BEC are made up of vortex
lines and soliton planes in the individual spinor components. The
construction of the interface-crossing defect solutions was achieved by
identifying defects in the two phases that have a similar combination
of vortex lines, and using soliton planes to achieve the switch
between polar and FM manifolds. Dark solitons have been
phase imprinted
experimentally~\cite{burger_prl_1999,denschlag_science_2000}.
Phase imprinting of vortex lines in a BEC by
transferring angular momentum from an electromagnetic
field has been proposed
theoretically~\cite{marzlin_prl_1997,bolda_pla_1998,williams_nature_1999,isoshima_pra_2000,dutton_prl_2004}
and several of the techniques have also been realized in
experiments~\cite{matthews_prl_1999,leanhardt_prl_2002,shin_prl_2004,andersen_prl_2006}.
In spinor BECs, coreless vortices and related textures have been
prepared by adiabatic 
ramping of a magnetic field along the trap
axis~\cite{leanhardt_prl_2003,choi_prl_2012,choi_njp_2012}, or by
population transfer using Laguerre-Gaussian laser~\cite{leslie_prl_2009}.
More 
complicated vortices and textures could be 
imprinted using proposed techniques for creating vortex
rings~\cite{ruostekoski_prl_2001,ruostekoski_pra_2005}.

Other interface defect configurations can be
constructed by combining more
elementary defect connections. For instance, in the FM phase the Dirac monopole
can be turned inside out to form an antimonopole, in which the spins
point radially inward.  By joining the Dirac strings of a Dirac
monopole and an antimonopole, a Dirac dipole can be
constructed~\cite{savage_dirac_2003}.  In
Fig.~\ref{fig:schematic2}(c), we illustrate a state where such a
dipole is placed on the polar-FM interface.  The Dirac string
forms a doubly quantized vortex line in the FM phase, connecting the
two monopoles.  Consequently, the Dirac
dipole can form the termination points of two oppositely winding singly
quantized vortices [see the construction that lead to
Eq.~(\ref{eq:s-dirac})].

\section{Core structure of interface-crossing defects}
\label{sec:core-structures}

\subsection{Core deformation of interface-crossing defect solutions}

In the preceding section we constructed the prototype spinor wave
functions for the
interface-crossing defect solutions connecting defects in the FM and
polar phases. We showed that such solutions can be formed by
combinations of elementary
vortex lines and dark soliton (phase kink) planes that could be experimentally
prepared by phase imprinting. Here we use the constructed spinor wave
functions for the defect configurations as initial states for
numerical studies of the defect
stability. Provided that prototype spinors sufficiently closely approximate the
local energetic minimum configuration, the initial states quickly
relax to the targeted defect structure.

By numerical simulations we can determine the energetically preferred
core structures and the energetic
stability of the defects. In order to do so we minimize
the free energy in the rotating frame [Eq.~(\ref{eq:free-energy})] by
propagating the coupled Gross-Pitaevskii equations, derived from
Eq.~(\ref{eq:hamiltonian-density}), in imaginary time using a
split-step algorithm~\cite{javanainen_jpa_2006}.  We assume the
slightly elongated trap, defined by
Eq.~(\ref{eq:V}). The initial state prototype spinor wave functions
are given in
Sec.~\ref{sec:defect-solutions}.  We choose
the spin-independent nonlinearity $c_0=2.0\times10^4\hbar\omega l^3$,
where $l=(\hbar/m\omega)^{1/2}$ is the transverse oscillator
length.  For $^{87}$Rb in a trap with $\omega=2\pi\times10$~Hz these
parameters correspond to $10^6$ atoms.  The spin-dependent
nonlinearity, $c_2$,  is allowed to vary.

In Ref.~\cite{lovegrove_pra_2012} it was demonstrated that the core of
a singly quantized vortex in the FM phase of spin-1 BEC deforms by
locally rotating
the spin vector so that the vortex lines in the individual spinor
components in the appropriate basis representation move apart.
The singular vortex line then no longer represents a vanishing atom density,
but is occupied by atoms with zero spin magnitude as in the polar
phase of the spin-1 BEC. The FM vortex line singularity filled by atoms
in the polar phase
becomes energetically favorable by allowing a larger core size and a
correspondingly smaller bending energy.  The core deformation can be
understood from
the energetics associated with the hierarchy of characteristic length
scales determined by the interaction strengths: the size of the filled
core is determined by the spin healing length $\xi_F$,
Eq.~(\ref{eq:xi_F}), which is usually larger than the density healing
length $\xi_n$, Eq.~(\ref{eq:xi_n}), that sets the size of a core with
vanishing density.

Similarly, a singly quantized vortex in the polar phase was
shown to lower its energy by spontaneously breaking axial symmetry, splitting
into a pair of singular half-quantum
vortices~\cite{lovegrove_pra_2012}.  This again avoids 
depleting the atom density in the vortex core:
at the location of the
singularities $\absF=1$, with spins anti-aligning in the two cores. The
two vortices form an extended core 
region where the order parameter is excited out of the polar
ground-state manifold.  The size of the core region is then enlarged
to be on the order of $\xi_F$, with a corresponding decrease in
bending energy.  
The overall topology is preserved away from the
two singularities. 
Inside the extended core region, the splitting of the
singly quantized vortex locally deforms the nematic $\nematic$ field that
describes the order parameter [see Eq.~(\ref{eq:nematic})],
and a disclination plane, where the identification $\nematic
\leftrightarrow -\nematic$ is made, appears between the vortex
lines. Thus on a loop encircling only one line singularity, both
condensate phase and nematic axis wind by $\pi$.  The splitting of the
singly quantized vortex is closely related to the deformation of a
point defect into a half-quantum vortex
ring~\cite{ruostekoski_monopole_2003}. In the 2D cross section of the
ring, the diametrically opposite
points on the ring correspond to half-quantum vortices with
anti-aligned spins in the cores. In Ref.~\cite{ji_prl_2008}
dynamic nucleation of half-quantum vortices under energy dissipation
was studied, demonstrating formation of a square vortex lattice.
To get a simple qualitative picture of
the interactions between half-quantum vortices 
one may consider 
the corresponding problem in a two-component BEC. In a nonrotating
uniform system it was argued that the repulsive force between vortices
with opposite 
core polarisations falls off as $1/R^3$~\cite{eto_pra_2011}.

\begin{figure}[tb]
  \centering
  \includegraphics[width=0.8\columnwidth]{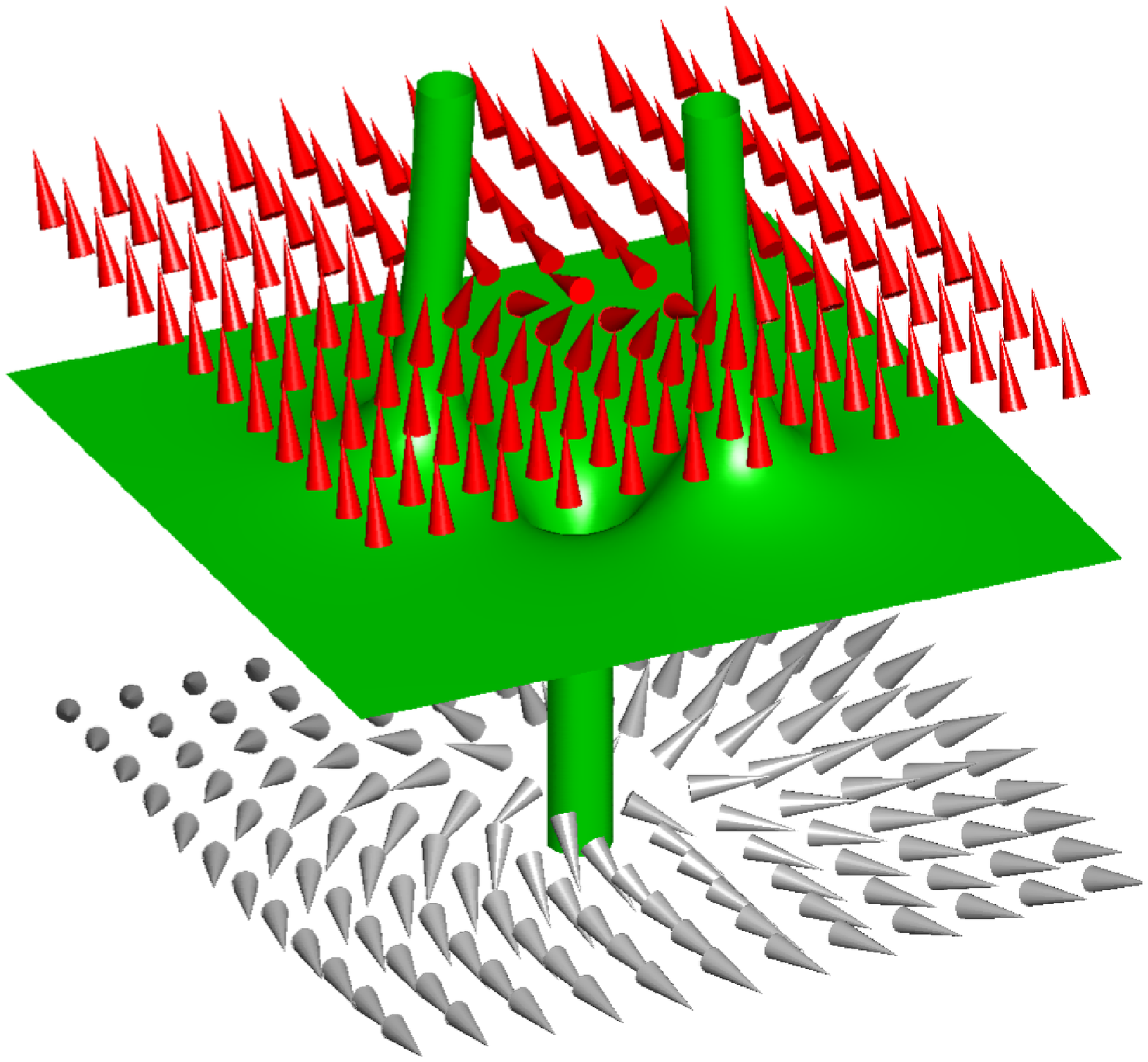}
\caption{(Color online) Deformation of connecting singly quantized
  vortices. The green (gray) $\absF$ isosurface shows the interface
  and the cores of singular
  defects.  The polar phase is above the interface. Energy relaxation
  causes the singly
  quantized vortex on the polar side to split into a pair of
  half-quantum vortices with $\absF>0$ cores on
  the polar side, recognizable by the characteristic $\pi$ winding of
  $\nematic$ (red/dark gray cones).
  These connect across the interface to a singly
  quantized FM vortex [cf.\ Fig.~\ref{fig:schematic2}(b)]. A local
  rotation of the spin vector (light gray cones) allows the core
  region to fill with atoms with $\absF<1$.
  ($\abs{c_2}=1.0\times10^4\hbar\omega l^{3}$ and $\Omega=0.20\omega$.)
}
\label{fig:111}
\end{figure}

It was demonstrated in Eq.~(\ref{eq:1qs}) how singly quantized
vortices in the FM and
polar phases can be connected across the polar-FM
interface, despite the fact that these are two topologically different
defects. As the energy is minimized, this interface-crossing
defect deforms by a mechanism analogous to that described above for the
singly quantized vortices in the purely FM and polar BECs. The
resulting structure is shown in Fig.~\ref{fig:111}.
On the polar side of the interface, the splitting of the singly
quantized vortex into two half-quantum vortices is recognized from the
deformation of the nematic field, which shows the characteristic $\pi$
winding around each singularity, and the formation of the disclination
plane.  The order-parameter is excited out of the $\absF=0$
ground-state manifold, to reach $\absF=1$ at the singular lines. 
The pair of
half-quantum vortices connects across the interface to the singly
quantized FM vortex.  This, in turn, exhibits the local rotation of
the spin vector, allowing the core region to fill by mixing FM and
polar phases, with $\absF=0$ on the singularity.

The relaxed
interface-crossing vortex structure is thus recognized as that
illustrated schematically in Fig.~\ref{fig:schematic2}(b), and the
deformation is understood in terms of the characteristic length scales
set by the atom-atom interactions.
In the purely polar or FM condensate, the core-deformed,
singly quantized vortices are energetically stable~\cite{lovegrove_pra_2012}.
For the parameter values investigated, the
configuration in Fig.~\ref{fig:111} ultimately decays for very long
relaxation times in our simulations.

\begin{figure}[tbp]
  \centering
  \includegraphics[scale=1]{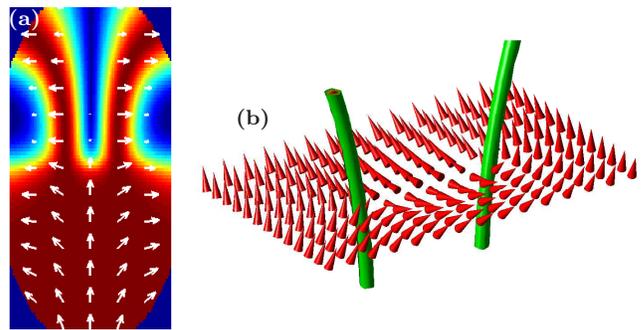}
\caption{(Color online) Minimizing the energy of Eq.~(\ref{eq:s-cl}),
  corresponding to Fig.~\ref{fig:schematic1}(b) results in a splitting
  of the singly quantized polar vortex, while the characteristic
  fountain-like spin structure of the coreless vortex is established in
  the FM part of the cloud.
  (a) The magnitude of the spin ($\absF=1$ is
  dark red with long arrows), shows the interface after relaxation of
  the energy, and 
  the filled cores, with $\absF=1$ at the singularity, of two
  half-quantum vortices in the
  polar part. White arrows show the spin vector and
  indicate the coreless vortex in the FM
  part, and how the spin structure connects to the vortex cores across
  the interface.  This result was obtained using
  $\abs{c_2}=2.5\times10^2\hbar\omega l^{3}$ and
  $\Omega=0.12\omega$.
  (b) The half-quantum vortices may be identified by the winding of
  the nematic axis $\nematic$ (unoriented but shown as cones to
  emphasize winding).  This displays the characteristic $\nematic \to
  -\nematic$ winding as any single vortex core is encircled.
  The two cores are joined by a disclination plane.  Note that away
  from the core region the original topology of the singly quantized
  vortex is preserved.
  Here a stronger spin-dependent nonlinearity has been used to get
  more sharply defined FM cores:
  $\abs{c_2}=1.0\times10^4\hbar\omega l^{3}$, $\Omega=0.19\omega$.
}
\label{fig:012}
\end{figure}

We do, however, find an energetically stable deformation
of a singly quantized polar vortex connecting across the interface if
instead of starting from $\zeta^{\rm 1\leftrightarrow s}$ of
Eq~(\ref{eq:1qs}), we minimize the energy of
$\zeta^{\rm 1\leftrightarrow cl}$ from Eq.~(\ref{eq:s-cl}).  This spinor
describes
a singly quantized polar vortex connecting across the interface to a
coreless vortex.  [A topologically equivalent configuration can be
constructed by allowing the polar vortex to terminate on a Dirac
monopole, Eq.~(\ref{eq:s-dirac}).]
Minimizing the energy leads to the
deformation shown in Fig.~\ref{fig:012}.  On the FM side of the
interface, the spin structure acquires the fountain-like structure
characteristic of the coreless vortex, as shown by the white arrows in
Fig.~\ref{fig:012}(a).  Here the frequency of rotation determines the
direction of the spin vector at the edge of the cloud as the
angular momentum in the FM phase adapts to the imposed rotation.

On the polar side of the interface we recognize the splitting of the
singly quantized vortex into a pair of half-quantum vortices,
identified by the $\pi$ winding of the nematic axis $\nematic$ around
each vortex [Fig.~\ref{fig:012}(b)],
preserving the overall topology. 
As before, in the core region the order parameter is excited out of
the ground-state manifold, with $\absF=1$ on the singular lines.
Fig.~\ref{fig:012}(a) shows how the spin texture
connects smoothly across the interface.

\begin{figure}[tb]
  \centering
  \includegraphics[scale=1]{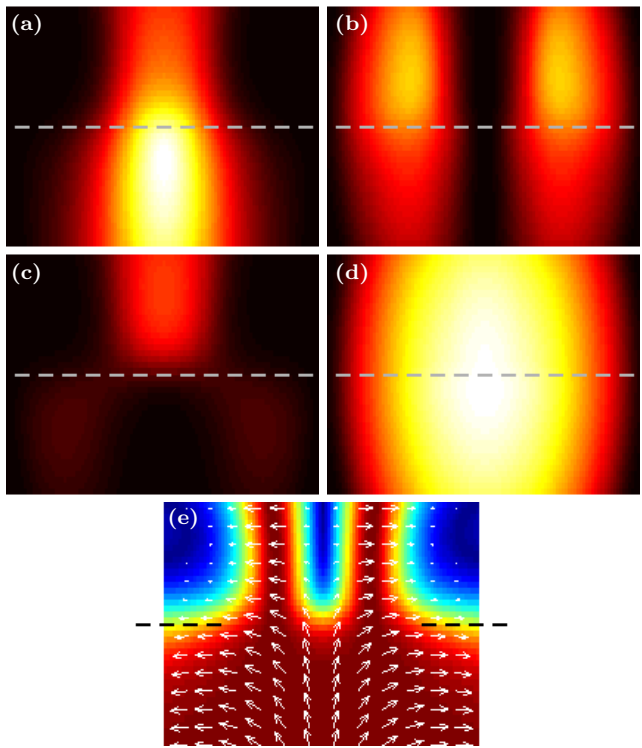}
\caption{(Color online) The condensate wave function varies smoothly across the
  polar-FM interface, showing that the defect states cross the
  interface smoothly, connecting the two topologies.
  (a)--(c) Densities in the individual spinor components
  $\zeta_+$,$\zeta_0$ and $\zeta_-$, respectively. These vary smoothly
  and do not vanish
  simultaneously on the interface.  The position of the interface,
  defined as the plane where $c_2$ changes sign, is
  indicated by the dashed line.
  (d) The smooth variation of the nonzero atomic density across the
  interface shows the continuity of the condensate wave function.
  (e) $\inleva{\mathbf{\hat{F}}}$ shown as color map (dark red with
  long arrows at
  $\absF=1$) and vector field (white arrows).  Note how the spin
  structure connects smoothly across the interface, in particular
  connecting the FM cores of the half-quantum vortices continuously to
  the fountain structure of the spin in the coreless vortex.  The
  magnitude of the spin also shows how the interface has acquired a
  finite width.
}
\label{fig:continuity}
\end{figure}

The continuity of the relaxed
spinor wave function is further demonstrated in
Fig.~\ref{fig:continuity}, giving a detailed picture of the interface
region.  In the relaxed state, the total atom density remains
nonvanishing at the
interface and varies smoothly across it.  The populations of the individual
spinor components are also continuous across the interface. The
continuity of the defect-carrying spinor wave function as it crosses
between the different broken symmetries means that it represents a
continuous connection of defects across the interface.
In Fig~\ref{fig:continuity}, the
position where
$c_2$ changes sign is indicated by a dashed line, and $\absF$ shown in
panel (e) shows the finite width of the interface region after energy
relaxation.

The vortex core structures are particularly intriguing and complex
when a vortex terminates to a point defect on the interface.
We study a singular FM vortex terminating to
a radial hedgehog at the interface ($\nematic$
forms a hemispherical hedgehog on the polar side), as depicted
schematically in Fig.~\ref{fig:schematic1}(d).  Unlike the FM Dirac
monopole, the polar monopole cannot unwind into simpler vortex
configuration.  Therefore, while the connection depicted in
Fig.~\ref{fig:schematic1}(b) is equivalent to
Fig.~\ref{fig:schematic1}(c) and leads to the same energy-minimizing
defect configuration, Fig.~\ref{fig:schematic1}(d) is topologically
distinct from all other vortex connections.

\begin{figure*}[tbp]
  \centering
  \includegraphics[width=\textwidth]{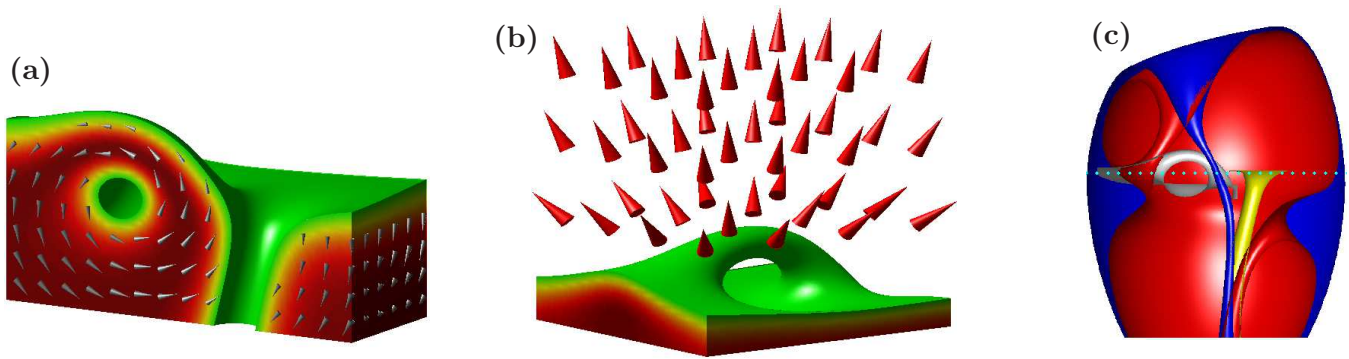}
\caption{(Color online)  In Fig.~\ref{fig:schematic1}(d) a polar
  point defect forms the termination of a FM spin vortex.  As the
  energy relaxes, the point defect deforms into a an arch-like,
  singular defect line that connects to the interface at both ends.
  (a) An isosurface of the spin
  magnitude is shown in green (light gray).  The spin magnitude rises
  to $1$ (dark red/black) on the FM side of the interface ($z<0$) and
  inside the line defect. Gray cones indicate the spin vector.
  (b) Away from the defect, the $\nematic$ vector field (red cones) in
  the polar
  phase retains the hedgehog structure of the original monopole.
  (c) In the trial wave function, Eq.~(\ref{eq:sv-pm}), both the FM
  spin vortex and the polar monopole are formed by exactly overlapping
  vortex lines of opposite winding in
  $\zeta^{\rm sv \leftrightarrow pm}_+$ and
  $\zeta^{\rm sv \leftrightarrow pm}_-$.
  Constant-density surfaces for $n\abs{\zeta_+}^2$ (red/medium
  gray) and $n\abs{\zeta_-}^2$ (blue/dark gray) show how the
  arch-shaped line defect is formed by deformation of these vortex
  lines, such that they no longer overlap close to the interface.
  The half-quantum vortex line (above the interface) and the
  spin vortex (below the interface)
  are indicated by silver and gold (light gray) spin
  isosurfaces at $\absF=0.9$ and $\absF=0.5$,
  respectively.
  ($\abs{c_2}=5.0\times10^2\hbar\omega l^{3}$ and $\Omega=0$.)}
\label{fig:alice}
\end{figure*}

\begin{figure}[tbp]
  \centering
  \includegraphics[width=\columnwidth]{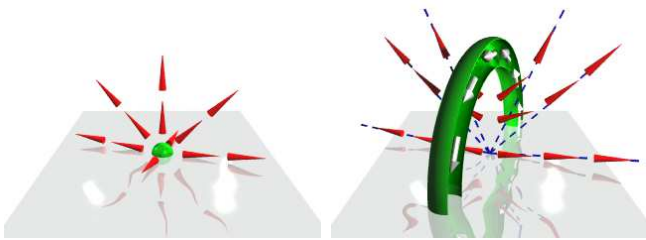}
\caption{(Color online).  Deformation of the hedgehog point defect (left)
  on the interface into an arch-shaped line defect (right). Red cones
  represent the
  (unoriented) $\nematic$ field; the defect cores are shown as a green
  sphere and torus, respectively. The deformation creates a hole in the
  point-defect core, allowing it to expand into the line defect by
  locally deforming the $\nematic$ field.  Away from the line defect,
  the asymptotic hedgehog structure is preserved. (Dashed lines indicate
  unperturbed $\nematic$ field for reference.)  On
  any closed loop through the arch, $\nematic$ therefore winds to
  $-\nematic$, identifying the defect as a half-quantum vortex
  line. The line-defect core is filled with atoms with  $\absF>0$;
  spin vector indicated by silver arrows.
}
\label{fig:deformation-schematic}
\end{figure}

A constructed prototype spinor wave function representing a singular FM vortex
terminating to a hedgehog point defect is given by Eq.~(\ref{eq:sv-pm}).
At the point defect singularity the atom density is zero.
The density depletion at the defect core is energetically costly, and
if $\xi_F$, Eq.~(\ref{eq:xi_F}), is sufficiently large in comparison
with $\xi_n$, Eq.~(\ref{eq:xi_n}), the
energy cost can be reduced by deforming the point defect into a
semi-circular line defect whose ends attach to the interface.  The
resulting arch-like defect is shown in Fig.~\ref{fig:alice} together
with the spin structure in the FM core (a) and the nematic axis away
from the defect on the polar side of the interface (b).  The
deformation of the defect, schematically illustrated in
Fig.~\ref{fig:deformation-schematic}, is
local and the topological charge of the
monopole is retained: away from the defect, the radial hedgehog structure of
$\nematic$ is preserved.  This implies that on any closed loop through
the arch, $\nematic$ must turn by $\pi$.  Consequently,
single-valuedness of the spinor wave function requires the condensate
phase to also turn by $\pi$, and we infer that the arch-shaped line
defect is a half-quantum vortex. This {\em Alice arch} resembles
the upper hemispheric
part of the Alice ring---a closed half-quantum vortex ring that exhibits the
topological charge of a point defect over any surface enclosing the
defect~\cite{ruostekoski_monopole_2003}.
Alice rings also appear in high energy physics~\cite{schwarz_npb_1982}
with a topological charge
similar to the magnetic ``Cheshire" charge~\cite{alford_npb_1991}.

The deformation of the point defect at the interface to an Alice arch results
from a complex interplay between the energetic considerations,
topology, and the length scale hierarchy in the system. The two
characteristic healing lengths determine whether the energy cost of
forming the extended vortex core region, where the singularity is
occupied by the atoms in the FM phase, offsets the energy cost of the
density depletion at a point defect (with a vanishing density at the
singularity). We find an abrupt transition
point to an Alice arch when $c_2 \lesssim 0.5c_0$. For larger values
of $c_2$ the point defect at the interface is preferred to the arch
defect. The sharp threshold for the deformation can be understood by
the topology. For the point defect there cannot be a singular point with
$\absF=1$ and a spherically symmetric core region with $\absF>0$. The nematic
axis forms a radial hedgehog structure and this would then imply that the spin
vector (always orthogonal to $\hat{\bf d}$) would have to form a
continuous tangent vector field for the spherically symmetric object
which is not possible (``hairy ball" theorem). Therefore the point
defect has to deform to a ring or an arch-shaped defect at the
interface before the 
transition from the zero-density singularity to the $\absF=1$
singularity with nonvanishing density is possible. The sharp
deformation threshold also appears 
in the stability analysis of a singular point defect and the Alice
ring~\cite{ruostekoski_monopole_2003}

We note that the arch defect is unstable towards drifting out of the
cloud due to the density
gradient resulting from the harmonic trapping potential.  It could
potentially be stabilized using a weak pinning laser to create a small
density depletion at the center of the trap.

The structure of the arch defect emerging from the point defect may
be understood by studying the individual spinor wave function components.
In the point defect with vanishing density at the singularity,
the overlapping vortex lines in $\zeta_\pm$ intersect with the soliton plane in
$\zeta_0$.  In the prototype spinor wave function
$\zeta^{\rm sv\leftrightarrow pm}$ an additional soliton plane is
present in $\zeta_+$ to account for the switch from the polar to the
FM side of the interface.  In Ref.~\cite{ruostekoski_monopole_2003}
the deformation of the spherically symmetric point defect was
explained by a local separation of the vortex lines in $\zeta_\pm$
such that they no longer overlap at the soliton plane.  Here the additional
soliton plane in $\zeta_+$ cuts the vortex line in
$\zeta_+$ at the interface, separating the two parts.
The spinor components $\zeta_\pm$ are shown in Fig.~\ref{fig:alice}(c).
The positions of the Alice
arch and the FM vortex are indicated, and we find how in the
polar part of the cloud, the separated vortex lines make up the
semi-circular half-quantum vortex. On the FM side, the unbroken vortex
line in $\zeta_-$ and the $z<0$ part of the vortex line in $\zeta_+$
form the spinor wave function of the FM vortex whose core is filled by
the vortex-free $\zeta_0$ component.

\begin{figure}[tb]
  \centering
  \includegraphics[width=0.98\columnwidth]{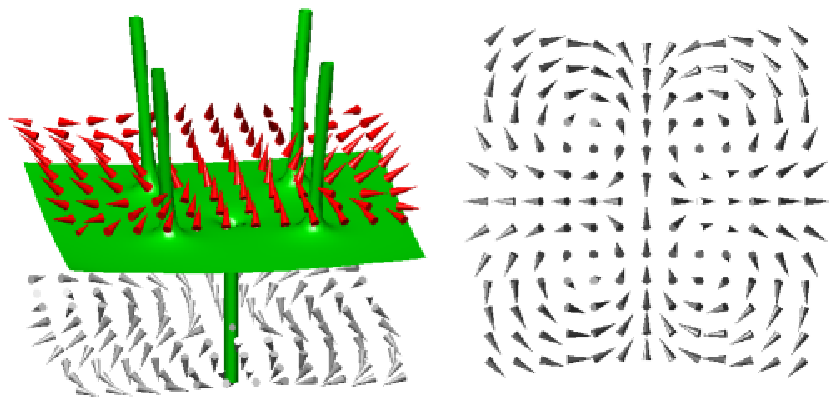}
\caption{(Color online) Left: Isourface of $\absF$ showing the cores
  of singular defects, together with $\nematic$ (red/dark gray
  cones) on the polar side of the interface, and
  $\inleva{\bvec{\hat{F}}}$
  (light gray cones) in the FM part.  Four nucleated vortices are
  identified as polar half-quantum vortices connecting across the interface
  to nonsingular coreless vortices. Right: Top-down view of
  $\inleva{\bvec{\hat{F}}}$ in the FM part, showing the four coreless
  vortices.
}
\label{fig:nucleation}
\end{figure}

At high rotation frequencies of the trap, we find nucleation of
interface-crossing defects in the energy minimization. Provided that
the appropriate instability for nucleating vortices is triggered, the
emergence of defect configurations  
where a half-quantum vortex connects to a coreless vortex
spontaneously emerges, as they lower the energy of the system in a
sufficiently rapidly rotating trap. An example is shown in
Fig.~\ref{fig:nucleation}, in which four interface-crossing 
vortices nucleate.
On the polar side of the interface, four singular lines appear, on
which $\absF=1$.  These vortices may be identified as half-quantum
vortices through the winding of the nematic axis.  On the FM side, the
order parameter remains nonsingular.  However, the spin texture
reveals that each of the four half-quantum vortices connects to a
coreless vortex.

\section{Conclusions}
\label{sec:conclusions}

Interfaces between topologically distinct ground-state manifolds play
an important role in several areas of physics, many of which are
difficult or impossible to access experimentally.
Here we have given a detailed analysis of our proposal for how
topological interface physics of defects and textures can be
studied in
ultracold atomic gases~\cite{borgh_prl_2012}.  We considered the
specific example of a spin-1 BEC
with spatially separated polar and FM regions within a continuous
condensate wave function.
For a number of representative,
interface-crossing defect states, we provided detailed constructions
of prototype spinor wave functions by considering how basic vortex and monopole
solutions with similar spinor structure represent different
topological objects in the two phases.  The resulting wave functions are built
from vortex lines and soliton planes in the individual spinor
components that could be phase-imprinted using existing experimental
techniques~\cite{matthews_prl_1999,leanhardt_prl_2002,shin_prl_2004,andersen_prl_2006,ruostekoski_prl_2001,ruostekoski_pra_2005,burger_prl_1999,denschlag_science_2000}.

The energetic stability and energy-minimizing core structures of the
interface-crossing defect configurations were analyzed by numerically
minimizing the energy of the prototype wave functions.
The resulting spinor states
demonstrate how the condensate wave function smoothly
interpolates between the two ground-state manifolds by locally
restoring its full
symmetry, thereby establishing a coherent interface through which defects may
connect continuously.
In particular, we demonstrated
the energetically stable connection of a coreless vortex to
a pair of half-quantum vortices, and the formation of an Alice arch:
the deformation of a point defect at the interface into an arch-shaped
half-quantum vortex line that preserves the topological charge.

In order to demonstrate the basic principle of the topological
interface physics in ultracold 
atoms, we have concentrated in this work on a relatively simple and
accessible example of defect perforation across constructed interfaces
in spin-1 BECs. 
The interface analysis, however, can also be applied to more complex
systems, such as  
\mbox{spin-2}~\cite{koashi_prl_2000,ciobanu_pra_2000,semenoff_prl_2007}
and
\mbox{spin-3}~\cite{barnett_prl_2006,santos_spin-3_2006} BECs, where,
for example, non-Abelian defects are
predicted~\cite{kobayashi_prl_2009,huhtamaki_pra_2009}. 
Other particularly promising platforms for topological interface studies are
strongly correlated atoms in optical
lattices~\cite{greiner_nature_2002,jordens_nature_2008,schneider_science_2008}
exhibiting also quantum phase transitions and potential analogues of exotic
superconductivity~\cite{bert_nphys_2011} in crystal lattices.

Moreover, the interface scheme may
be used to investigate nonequilibrium dynamical scenarios for 
production of topological defects and textures in phase transitions. 
An intriguing possibility is production of topological defects in
experiments inspired by brane-inflation models, where brane
annihilation leads to formation of
defects~\cite{dvali_plb_1999,sarangi_plb_2002}.  In a FM BEC, a region
of polar phase can be created by locally shifting the spin-dependent
interaction strength. The resulting phase boundaries then form
two-dimensional analogues of $D$-branes.  When the interaction shift
is removed, the polar region collapses, simulating brane-antibrane
annihilation, and resulting defects can be observed in the cloud.
This is similar to a recent experiment in
$^3$He~\cite{bradley_nphys_2008}, where,
however the defects are more difficult to observe directly.

Defect production may also result from dynamic instabilities,
for example arising from superfluid counterflow between the FM and
polar phases.
The boundary between two fluids moving with respect to each other
becomes unstable if the relative velocity exceeds some critical value,
leading to excitations on the interface.  This
phenomenon is well understood in classical fluid mechanics and is known
as the Kelvin-Helmholtz instability~\cite{kundu-cohen}. An analogous
\emph{superfluid} Kelvin-Helmholtz instability has been shown to
occur at the interface between superfluid $^3$He $A$ and
$B$, providing another active area of research related to interfaces
between different ordered phases~\cite{finne_rpp_2006}.
Vortices in the $A$ phase cause counterflow against the initially
vortex-free $B$ phase.  As the relative velocity exceeds a critical
value, vortices nucleate from the interface into the $B$ phase.
A superfluid Kelvin-Helmholtz instability has also been predicted in
phase-separated two-component
BECs~\cite{suzuki_pra_2010,lundh_pra_2012}.

\acknowledgments
We thank D.~J.~Papoular for enlightening
discussions. Financial support from the Leverhulme trust is gratefully
acknowledged.

\end{document}